\documentclass[prl,twocolumn,floats,aps,superscriptaddress,showpacs]{revtex4}

\usepackage{amsmath}
\usepackage{epsfig}

\newcommand{\bx}{{\bf  x}}

\newcommand{\br}{{\bf  r}}

\newcommand{\bk}{{\bf  k}}

\newcommand{\BE}{\begin{equation}}
\newcommand{\EE}{\end{equation}}
\newcommand{\BA}{\begin{array}}
\newcommand{\EA}{\end{array}}
\newcommand{\beqn}{\begin{eqnarray}}
\newcommand{\eeqn}{\end{eqnarray}}

\newcommand{\etab}{\mbox{\boldmath $\eta$}}

\newcommand{\nablab}{\mbox{\boldmath $\nabla$}}

\begin{document}

\title{Macroscopic description of particle systems with
non-local density-dependent diffusivity.
}

\author{Crist\'obal L\'opez}
\affiliation{Instituto
Mediterr\'aneo de Estudios Avanzados IMEDEA (CSIC-UIB),
     Campus de la Universidad de las Islas  Baleares,
E-07122 Palma de Mallorca, Spain.
} 

\date{\today}

\begin{abstract}

In this paper we study macroscopic
density equations
in which the diffusion coefficient
depends on a weighted spatial average of the density itself.
We show that large differences
(not present in the local density-dependence case)
appear between
the density equations that are derived from
different representations
of the  Langevin
equation describing a  system of interacting
Brownian particles.
Linear stability analysis
demonstrates that  under some circumstances the  density
equation interpreted like Ito
has pattern solutions, which
 never appear for the 
H\"anggi-Klimontovich interpretation, which is the other one typically
appearing in the context of nonlinear diffusion processes.
We also introduce a discrete-time microscopic model of 
particles that
confirms the results
obtained at the macroscopic density level.

\end{abstract}
\maketitle



In the mathematical description of
many ecological systems the mobility of the species 
depends on the local density or concentration of the individuals \cite{okubo,murray}. 
A standard example of this is given by models for animal or insect
dispersal where it is usually considered that there is an
increase of the diffusion coefficient due to population pressure. 
These models are written down as
an extension of the diffusion equation 
which includes 
a diffusivity depending on the population density.
Under the generic name of nonlinear diffusion equations
similar  equations arise in other contexts like in 
the flow through a porous medium, bacterial dynamics,
 and  transport in 
plasmas~\cite{murray,aronson}.

Generally, the nonlinear diffusion
equations for the density
correspond to 
a coarse-grained description of the more fundamental 
particle stochastic dynamics. Similarly to  the case of diffusion
in nonhomogeneous media where the diffusion coefficient
varies in space, a heuristic derivation 
of the nonlinear diffusion density equations 
has to deal with the Ito-Stratonovich
dilemma \cite{vankampen, sancho}. (To be  precise
one should speak about the interpretation of 
the Langevin equation  since,
as we will see later, the {\it Stratonovich}
realization does not appear in our study
and one has instead the H\"anggi-Klimontovich
interpretation of the Langevin equation \cite{hanggi}.)
In the context of noise-induced
phase transitions this is of a major importance
\cite{raul},
and this work pretends to show that it 
is also specially relevant when one considers 
nonlocal interactions between  the particles.

In a biological setting, 
nonlocal effects have been widely considered
since they account for the interaction between
individuals that are separated in space
\cite{flierl}.
More recently, there has appeared  interest in models
where the mobility depends on an 
averaged density rather than on its
value at a point \cite{holm}.  
The spatial non-locality is  there
introduced in order
to take into account  the
finite-size of the particles or 
the senses (visual, hearing, etc..) of the organisms, that
somehow determine their mobility. 
The idea behind is that the diffusivity of any
particle may depend on the configuration 
of other particles in its vicinity.

Thus, 
in this work we study a nonlocal version of a general
density-dependent diffusion equation for
the number density of individuals.
We focus on its pattern and clustering formation
properties. 
Making linear stability analysis we first show that,
contrarily to what happens in the local case, completely
different results can be obtained for the
Ito and H\"anggi-Klimontovich
descriptions.
In fact, spatial patterns can emerge for the first but never for
the second.
Then, we introduce two microscopic    
interacting particles systems
whose continuum evolutions 
are given by the 
Ito and H\"anggi-Klimontovich density equations.
Numerical simulations of these systems 
confirm the calculations made
with the continuum density equations.

Let us consider a population dynamics
model of relevance for insect dispersal.
The diffusion coefficient depends on
the local density of particles, i.e., $D=D(\rho)$, where $\rho (\bx ,t)$ is
the population density (we assume that  $D$ is not an explicit function of  the 
spatiotemporal coordinates). 
We only study
positive density-dependence, which
is the most widespread hypothesis and indicates that 
competition increases dispersing because individuals have
better survival rates by leaving high-density areas~\cite{ecography}. 
Mathematically, $D'(\rho) \equiv dD/d\rho >0$. 
The mean-field density descriptions in Ito
and H\"anggi-Klimontovich interpretations
 are~\cite{vankampen,raul,hanggi}:
\begin{eqnarray}
\frac{\partial  \rho (\bx, t)}{\partial t}&=&\nablab^2 \left( D(\rho) \rho \right),
\label{IDESDE}
\\ 
\frac{\partial  \rho (\bx, t)}{\partial t}&=&\nablab \cdot \left( D(\rho) \nablab
\rho \right).
\label{SDESDE}
\end{eqnarray}
The first one corresponds to Ito and the second to H\"anggi-Klimontovich, and in the
following we will denote them, respectively, as IE and HKE.
The HKE is the one that is 
usually considered when modeling at the macroscopic level~\cite{okubo,murray},
since one writes down the density evolution as 
a flux equation $\partial_t \rho =- \nablab \cdot {\bf J} $
with the density-dependent flux ${\bf J}=-D(\rho)  \nablab \rho$.
But this is
not a definitive argument and, as it is
discussed in \cite{horstemke}, the IE 
should be the correct one  for a continuum description
of a discrete population dynamics model of non-overlapping generations. 
A proper derivation of density  equations from the fundamental
stochastic particle dynamics,  once a right interpretation
for this is
assumed, can be consulted in \cite{dean}. Note that they are stochastic,
but in our study we neglect the fluctuations by considering a
kind of mean-field approximation \cite{umberto}.
Let us
assume an initial 
constant density $\rho (\bx, 0) =\rho_0$. We now perform a linear
stability analysis of the stationary homogeneous (identical to $\rho_0$
because particle number is conserved) solution for 
both IE and HKE. Writing $\rho (\bx, t) =\rho_0 + \epsilon \psi (\bx, t)$,
where $\epsilon$ is a small parameter, and $\psi$ the space-time dependent
perturbation, we obtain (we label the equations with IE or HKE):
\begin{eqnarray}
(IE) \ \  \partial_t \psi (\bx, t)& =& \left( D\left(\rho_0\right)+ \rho_0 D' \left(\rho_0\right)\right)
\nablab^2 \psi \\
(HKE) \ \  \partial_t \psi (\bx, t) &=& D(\rho_0)
\nablab^2 \psi.
\label{locales}
\end{eqnarray} 
Note that in both the perturbation follows a simple diffusion equation and there
is, therefore, no instability in the density equation that could give rise
to the formation of patterns or aggregates of particles. From this point of view,
the Ito and H\"anggi-Klimontovich
frameworks (Eqs.~(\ref{IDESDE}) and (\ref{SDESDE})) are equivalent. 

Now let us consider that the diffusivity depends on an
averaged density over
the whole system, ${\bar \rho} (\bx, t)=\int_{\Re^d}
 d\br G(\bx -\br) \rho (\br, t)$,
where the kernel, $G(\bx -\br)$, 
represents the effect of population density at $\br$ on the density at $\bx$.
It embodies the forces of attraction and repulsion of
neighbors~\cite{kenkre}.
We assume that $G$ is normalized so that $\int_{\Re^ d} d\br G(\br )=1$.
Then let us consider $D=D({\bar \rho})$ in
Eqs.~(\ref{IDESDE}-\ref{SDESDE})
 and make again a
linear stability analysis around $\rho_0$
for the now nonlocal HKE and IE,
\begin{eqnarray}
(HKE) \nonumber \\ &\epsilon \partial_t \psi (\bx, t) = 
\nablab \left[ D(\rho_0 +\epsilon \int d\br G(\bx -\br) \psi (\br, t)) \right. & \nonumber \\
&\left. \nablab (\rho_0 + \epsilon \psi) \right],& 
\label{sdenolocal}
\end{eqnarray} 
so that to first order in $\epsilon $ the perturbation simply diffuses,
$\partial_t \psi = D(\rho_0)\nablab^2 \psi$, and no instability
happens in HKE. 
However, when we consider the Ito description 
we have:
\begin{eqnarray}
&(IE)\ \ \ \ \ \ \  \ \ \epsilon \partial_t \psi (\bx, t) \ \ \ \  & \nonumber \\
&=\nablab^2 \left[ D\left(\rho_0 +
\epsilon \int d\br G(\bx -\br) \psi (\br, t)\right) 
\left( \rho_0 + \epsilon \psi (\bx, t)\right) \right]& \nonumber \\
&= \epsilon \nablab^2 \left[ D(\rho_0) \psi + \rho_0 D(\rho_0)\int d\br 
G(\bx-\br) \psi(\br, t)\right].& 
\label{idenolocal}
\end{eqnarray}
Considering a harmonic perturbation 
$\psi (\bx, t)= \exp(\lambda t+ i \bk \cdot \bx)$
it is not difficult to obtain the following dispersion
relation
\begin{equation}
\lambda (k)= -k^2 \left( D(\rho_0)+
\rho_0 D'(\rho_0) {\hat G} (\bk) \right),
\label{dispersion}
\end{equation}
where $k=|\bk|$ and
${\hat G}(\bk)=\int d\br G(\br) e^{-i\bk\cdot \br}$ is the Fourier
transform of the kernel. 
The important fact with Eq.~(\ref{dispersion}) is that
depending on the kernel function  (note that the
form of the diffusivity functional is almost irrelevant 
once we have assumed that it is an increasing function
of the density, $D'(\rho)>0$),
the perturbation growth
rate can be positive, giving rise to the aggregation
of particles or formation of spatial patterns~\cite{flierl,kenkre}.
As we have
just seen, this is absolutely different to the result
obtained in the H\"anggi-Klimontovich
framework, Eq.~(\ref{sdenolocal}) and below. 

\begin{figure}
\begin{center}
\epsfig{figure=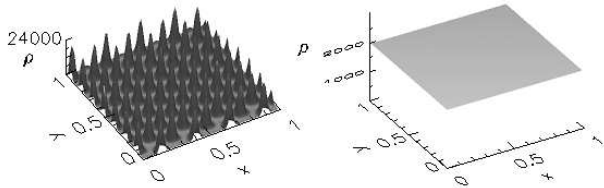,clip=true,width=\columnwidth,angle=0}
\end{center}
\caption{Steady number density of particles with nonlocal density-dependence
diffusivity.  
Left is Ito and right H\"anggi-Klimontovich.
The diffusivity functional is  $D(\rho)=D_0(\rho/\rho_0)^p$, and
we use the top-hat kernel with $R=0.1$. The values of the 
other
parameters, in both plots,
 are $D_0=0.0001$, $p=9$, $\rho_0=2000$, and system sizes
in $x$ and $y$  are equal to $1$.
}
\label{fig:patternscont}
\end{figure}

Next we introduce a model, choosing 
the density and the kernel functions,
with a clearly oriented
ecological application and where $\lambda$ can take positive values.
We study the distribution of organisms over a spatial
area and  thus we 
restrict to two spatial dimensions.
A typical form for
$D(\rho)$ in 
animal dispersal models is $D_0 (\rho/\rho_0)^p$ where $D_0$ and $p$ are
positive real numbers~\cite{okubo,murray}, and $\rho_0$ is a reference density
which we take equals to the initial one without loss of
generality. Because of the normalization of the
kernel function the averaged density, ${\bar \rho}$, has
the same  dimensions as $\rho$.
Concerning the kernel, we 
take the usual top-hat function \cite{kenkre}:
$G(\br)=1/(\pi R^2)$ if $|\br | \leq R$ and 
$G(\br)=0$ otherwise, which introduces a typical
interaction length, $R$, in the system.
One obtains
${\hat G}(k)=2 J_1(k R)/(k R)$ so that the dispersion relation
takes the form
\begin{equation}
\lambda (k)= - D_0 k^2 \left( 1 + \frac{2 p J_1 (kR)}{kR} \right).
\label{dispersion2}
\end{equation}
$J_1$ is the first order Bessel function, and it is clear from
Eq.~(\ref{dispersion2}) that
the value of $D_0$ only sets the time scale of the system. 
The onset of pattern formation is $\lambda >0$ which numerically 
is approached for $p$ larger than $p_c \approx 7.6$.
In Fig.~\ref{fig:patternscont}
we show a long-time density solution for the IE (left) and 
HKE (right) with this kernel and $D$ functions, and $p=9$.
Note the hexagonal pattern formed in IE and the homogenous solution for
HKE, i.e., 
the numerical results
confirm the different spatial structures obtained  for the
two descriptions. We have checked that this is so for
all $p$,  as can be seen  below in Fig.~\ref{fig:structure}.

Up to now we have shown that, for
nonlinear nonlocal diffusion phenomena,
the different 
interpretations, in terms of density equations,
of the same system dynamics lead to 
very distinct results.
One may  ask if this
also happens at the level of the discrete
interacting particle dynamics.
Since the IE and HKE
are the  mean-field density
descriptions 
derived from 
the Langevin equation
of a system
of Brownian interacting particles,   
we expect that
the same differences encountered before
between both descriptions
appear.
We will show  that this is in
fact the case for two time-discrete microscopic 
systems
which,  in the continuous time limit,
are the Ito and H\"anggi-Klimontovich
Langevin equations corresponding to the IE and HKE 
with the top-hat kernel and the typical diffusivity
functional for insect dispersal. 

Therefore, the
problem is to find a discrete dynamics  from
which  deriving HKE and IE. 
Following the  discussion
in \cite{horstemke} (Section 5.4.2)
we first take a population dynamics 
system of non-overlapping generations whose 
Langevin equation, in the time continuum limit,
should 
be interpreted in Ito. The system consists in  
$N$ particles with positions 
$\br_i (t)=(x_i (t), y_i (t))$, $i=1,..., N$, in a two-dimensional periodic box of 
size $L=1$
evolving as follows
\begin{equation}
\br_i(t+\Delta t)= \br_i(t) + \sqrt{2 D_0 (N_R (\br_i(t))/N)^p \Delta t}
\ \etab^{(i)} (t), 
\label{itodiscreta}
\end{equation}
where $D_0$ and $p$ are real positive numbers, $\Delta t$ is the time step,
$N_R (\br_i(t))$  is the number
of particles at a distance less than $R$ of particle $i$ at time $t$,
and
$\etab^{(i)}(t)=(\eta^{(i)}_x, \eta^{(i)}_y)$ 
is a Gaussian White noise with
correlations $\left<\eta^{(i)}_a (t) \eta^{(j)}_b (s) \right>=
\delta_{i j} \delta_{a b} \delta_{t s}$. 
As already mentioned, in the 
limit  $\Delta t \to 0$
one obtains an Ito Langevin equation,
\begin{equation}
\frac{d\br_i}{dt}=\sqrt{2 D_0 (N_R (\br_i(t))/N)^p} \etab^{(i)} (t),
\label{itole}
\end{equation}
and one can  calculate~\cite{dean,umberto}
that
the corresponding density equation of this Langevin is 
the nonlocal IE with the top-hat kernel
and the density functional of insect dispersal. 
For the H\"anggi-Klimontovich case, the first step is to interpret
eq.~(\ref{itole}) in this sense so that it is equivalent to the
following Ito Langevin equation \cite{nota}:
\begin{equation}
\frac{d\br_i}{dt}=g(\br_i(t))\nabla_{\br_i} g(\br_i(t))+
\sqrt{2 D_0 (N_R (\br_i(t))/N)^p} \etab^{(i)} (t),
\label{hkle}
\end{equation}
where $g(\br_i(t))=\sqrt{2 D_0 (N_R (\br_i(t))/N)^p}$. 

\begin{figure}
\begin{center}
\epsfig{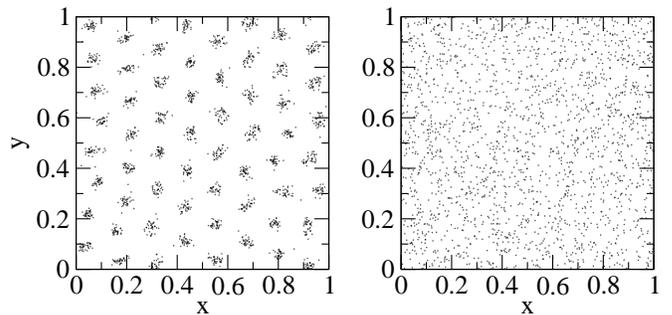}
\end{center}
\caption{Spatial distribution of particles at  large time. 
Left, is for  Ito  and right for  H\"anggi-Klimontovich.
We use the same parameters as in Fig.~(\ref{fig:patternscont}),
i.e., $D_0=0.0001$, $p=9$, and $R=0.1$. The number of particles is $N=2000$,
 for
both plots, and  $dt=0.01$.
 For the right panel we take a square grid (to calculate the gradients
of the $g$ function)  with lattice spacing $dx=0.005$.
}
\label{fig:patternsdis}
\end{figure}

Thus, our discrete particle systems equivalent to the IE and HKE
are given by Eq.~(\ref{itodiscreta}) and  the discrete-time version
up to order  $\Delta t$ of Eq.~(\ref{hkle}), respectively.
Note that, since  our simulations are off-lattice, i.e.,
particles move in the continuum space, 
we have to consider an auxilary
grid to compute the gradients of $g(\br_i(t))$
for the H\"anggi-Klimontovich case. 
In Fig.~(\ref{fig:patternsdis}) we show 
the long-time spatial distribution of particles
for the same parameter values used in Fig.~(\ref{fig:patternscont}). 
Left corresponds to  Ito
 and right to 
H\"anggi-Klimontovich.
The particles following the time-discrete-dynamics
which
give rise to the Ito Langevin equation arrange
in an hexagonal pattern of clusters, while in the
H\"anggi-Klimontovich case the particles are  randomly
distributed in the space. This is completely in accord
with what it happens in the macroscopic description.

A more quantitative comparison between
the macroscopic and microscopic dynamics follows. 
The characterization of the pattern 
can be performed via
the  structure function,  whose
maxima identify periodic structures. In the case of 
the continuum density, the structure function in the steady state, $S_c$, 
is computed as the modulus of the Fourier
transform of $\rho (\bx, t)$ averaged spherically
and in time. In the case of the particle system
the structure function, which is related but not
identical to $S_c$, is calculated as
$S_d(k)=\left< \left| \sum_j e^{i \bk \cdot \bx_j (t)}\right|^2 /N
\right>$, where 
$\bx_j (t)$ is the position vector of particle $j$, 
$\bk$ is a two-dimensional wave vector with modulus $k$,
and the average indicates  a spherical average over the wave vectors
with modulus $k$,  a temporal average in
the steady state, and an average
over many different realizations.
In Fig.~\ref{fig:structure}
we plot the  maximum (for $k>0$) value of $S_c$ (left panel) and of $S_d$ (right)
versus the parameter $p$. With circles we plot the computed ones
for IE and with squares for HKE.
It is clear from 
the plot that the value of $p$ at which the transition to pattern,
in Ito,
occurs is the same for both, and coincident with
the numerical value obtained from stability analysis.
Moreover, for any value of $p$ no peak (i.e., no pattern)
is ever observed in the H\"anggi-Klimontovich prescription.

\begin{figure}
\begin{center}
\epsfig{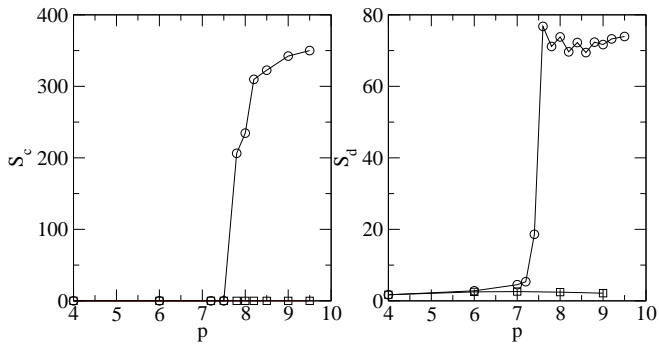}
\end{center}
\caption{Maxima of the structure functions for different values of
the parameter $p$. In the left panel we plot the maximum value of $S_c$,
i.e., the corresponding to the continuum density equation. It is normalized
with the number of points of the grid used to make the computation, $10201$.
Circles are for the Ito prescription and squares for the H\"anggi-Klimontovich one.
On the right, we plot, also vs $p$,
 the maximum value of $S_d$, i.e, of the structure function
for a  distribution of particles. Also circles are for Ito and squares
for H\"anggi-Klimontovich. The rest of the parameters for both plots are $D_0=0.0001$
and
$R=0.1$. For the H\"anggi-Klimontovich  particle dynamics we use
a lattice  with  $dx=0.005$.
}
\label{fig:structure}
\end{figure}

In summary,
density evolution equations with density-dependent  diffusion
coefficient
are standards in  the macroscopic 
modeling of biological, physical and
chemical processes. 
Typically, they  are postulated
as $\partial_t \rho (\bx, t)= - \nablab \cdot {\bf J}$
with ${\bf J}$ the population flux given by
${\bf J}= -D(\rho) \nablab \rho$. In this way one
obtains a macroscopic density equation in the
so-called
H\"anggi-Klimontovich framework, which is  different to the Ito
one, Eq.~(\ref{IDESDE}).
Though this last expression
rarely appears in the literature,
this is no reason to exclude it~\cite{vankampen},
and one has to resort
to the real microscopic dynamics  to infer the proper density
equation describing the system dynamics. 
In fact, a proper macroscopic equation for population
dynamics models of non-overlapping generations goes through Ito
\cite{horstemke}.
In this work we have
shown that,  in the context of biological population 
dynamics but easily extended to others,
no relevant differences, at the level of clustering or
pattern formation,  arise between
both equations when the functional $D(\rho)$ is local. 
However, when $D$ is  nonlocal 
the differences 
are rather important, and non-homogenous
spatial structures can appear in  Ito but never
in H\"anggi-Klimontovich. 
Furthermore, we have discussed the discrete 
dynamics of non-locally
interacting particle systems 
from which the Ito and H\"anggi-Klimontovich  density
equations are derived, confirming also at this level
the large differences encountered in the macroscopic
description. 
It is thus clear that the main message of this work
is that  it is specially relevant in the case
of systems with 
nonlocal density-dependent diffusivity  that
the macroscopic description should be properly derived
from the discrete particle model.
The inclusion of  reaction terms and further extensions 
of the model will be pursued 
in the future.

I acknowledge useful conversations with
Emilio Hern\'andez-Garc\'\i a and   Umberto
Marini Bettolo Marconi. 
Financial support from MEC (Spain) and FEDER through project
CONOCE2 (FIS2004-00953) is greatly acknowledged.
C.L. is a {\sl
Ram{\'o}n y Cajal} research fellow of the Spanish MEC.



\begin{thebibliography}{99}
\bibitem{nota}
According to the H\"anggi-Klimontovich interpretation, 
the discrete-time version of the Langevin equation 
Eq.~(\ref{itole})  goes through the
post-point discretization. However, in order to have
a dynamics of non-overlapping generations we 
write down  first its equivalent
Langevin equation in Ito, Eq.~(\ref{hkle}),
and then  consider its time-discrete form to
order  $\Delta t$.
\end{thebibliography}
\end{document}